\documentclass[11pt]{article}

\usepackage[utf8]{inputenc}
\usepackage[T1]{fontenc}
\usepackage[margin=1in]{geometry}
\usepackage{amsmath,amssymb}
\usepackage{graphicx}
\usepackage{booktabs}
\usepackage{caption}
\usepackage{float}
\usepackage{textcomp}

\usepackage[
  pdftitle={Statistically Significant Linear Alignments Among High-Confidence Transient Candidates on POSS-I Photographic Plates},
  pdfauthor={Brian Doherty},
  pdfsubject={Astronomy; POSS-I transient candidates; alignment statistics},
  pdfkeywords={methods: statistical, surveys, catalogs, POSS-I, VASCO},
  pdfcreator={Brian Doherty},
  pdfproducer={pdfTeX},
  hidelinks
]{hyperref}

\pdftrailerid{}
\title{Statistically Significant Linear Alignments Among High-Confidence Transient Candidates on POSS-I Photographic Plates}
\author{Brian Doherty\\
\small Independent Researcher, Dallas, TX, USA\\
\small \texttt{briandohertyresearch@gmail.com}}
\date{April 2026}

\begin{document}
\maketitle

\begin{abstract}
\noindent\textbf{I report the detection of statistically significant linear alignments and anomalous spatial clustering among high-confidence transient candidates in the VASCO catalog of vanishing sources on Palomar Observatory Sky Survey (POSS-I) photographic plates (1949--1957). A machine learning classifier scores 107,875 candidates by their likelihood of being genuine transients. Searching the 36,215 candidates with probability $\geq 0.50$ for collinear groupings narrower than 3 arcsec, I find 7 plates with alignments of 5--8 sources that exceed Monte Carlo expectations ($p < 0.03$, 10,000 iterations). The aligned sources are point-like, not streaks, which rules out any continuously luminous object crossing the field during the 45-minute exposures. The implied angular rates (1--15 arcsec s$^{-1}$) overlap with the geosynchronous regime (Blake et al., 2021) but are inconsistent with low or medium Earth orbits, and no artificial satellites existed during the POSS-I era. When I project each alignment onto Earth's surface assuming a high-altitude object, 6 of 7 maintain constant geographic longitude with sub-degree spread (combined $p \sim 3 \times 10^{-10}$). Four of these cluster near $-96^\circ$ longitude (central United States); one falls within $0.3^\circ$ of the longitude of the Hanford nuclear production site on a nuclear test window date. Close pairs ($< 30$ arcsec) occur at 16.2$\times$ the random rate, and the nights with alignments are the same nights with excess close pairs (Fisher exact $p < 0.0001$). Plate artifacts cluster near the ecliptic plane (26\%), but high-confidence transients are depleted there (16\%; $\chi^2$ test $p = 3.3 \times 10^{-82}$), which rules out asteroids, comets, and zodiacal debris as the dominant source. No transient reappears at the same sky position on a different night. All of these transients predate Sputnik 1.}
\end{abstract}


\section{Introduction}

The VASCO project (Villarroel et al., 2020) has cataloged 107,875 transient candidates on Palomar Observatory Sky Survey (POSS-I) plates taken between 1949 and 1957. These are sources that appear on a single plate exposure but have no counterpart in modern surveys or on temporally adjacent plates (Solano et al., 2022).

I previously trained a 23-feature ensemble machine learning classifier on 250 human-labeled candidates using morphometric measurements from red-band FITS cutouts (Bruehl, Doherty et al. 2026, arXiv:2604.18799). The classifier assigns each candidate a probability of being a genuine transient rather than a plate artifact.

Two findings from prior work motivate this investigation. Bruehl \& Villarroel (2025) showed that transient detection rates are elevated during nuclear test windows (+/-1 day from detonation), a result independently replicated in Doherty (2026b). Villarroel et al. (2025) showed that transients avoid Earth's geometric shadow at geosynchronous altitude, also confirmed in the replication. Villarroel et al. (2025) also reported alignments among VASCO candidates, but without pre-filtering the sample by likelihood of being a real transient versus a plate artifact. The present analysis is the first to test for alignments specifically among candidates that have all been independently scored as likely real by an ML classifier trained on morphometric features. The earlier findings point toward reflective objects in near-Earth space, but none of them address the spatial arrangement of high-confidence transients on individual plates. That is the question I take up here: do high-confidence transients exhibit alignments more often than chance, and if so, what does that geometry tell us?

There are several reasons why transients might line up on a plate. A satellite or debris fragment could produce intermittent glints as it crosses the field during a long exposure. Cosmic ray particle tracks can leave linear marks. Plate defects sometimes run along edges or emulsion boundaries. Or sources might align by chance. I develop a systematic search, test each result against Monte Carlo randomization, and then look at what the geometry implies.

\section{Data}

The input catalog is the VASCO v4 validated catalog (Solano et al., 2022; Villarroel et al., 2025), containing 107,875 candidates with J2000 equatorial coordinates, observation dates, and plate identifiers. ML probabilities come from the red all-candidates ensemble classifier (Bruehl, Doherty et al. 2026, arXiv:2604.18799), which scores every candidate using 23 morphometric features extracted from red-band FITS cutouts.

I restrict the alignment search to the 36,215 candidates with ML probability $\geq 0.50$, spread across 635 plates and 306 observing nights.

\subsection{Sky Coverage of the Parent Sample}\label{sec:skycov}

\begin{figure}[H]
\centering
\includegraphics[width=0.92\textwidth]{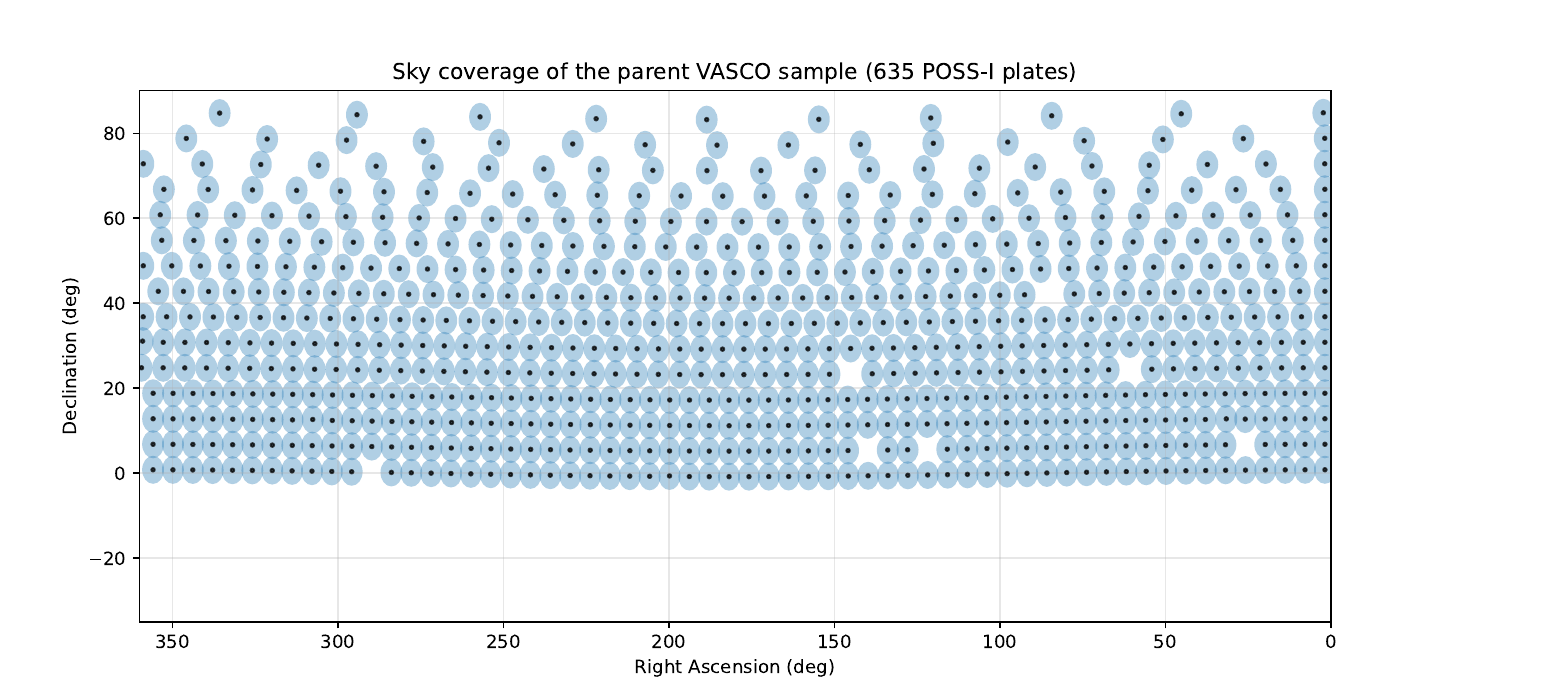}
\caption{Sky coverage of the 635-plate parent VASCO v4 sample. Each circle marks a POSS-I plate footprint at radius $3.3^\circ$. The sample tiles the northern sky uniformly across the full RA range, with no large empty regions.}
\label{fig:skycov}
\end{figure}

The plates used in this analysis are drawn from a complete and evenly distributed parent sample. Figure~\ref{fig:skycov} shows right ascension versus declination for all 635 POSS-I plates that contribute candidates to the VASCO v4 catalog (Solano et al., 2022). The parent sample tiles the northern sky from declination $0^\circ$ to $+90^\circ$ across the full RA range with no large coverage gaps. This is a property of the catalog's construction: VASCO v4 detects sources directly from the POSS-I red-band scans and does not pre-filter by cross-matching against external multi-wavelength catalogs. Catalogs that do apply such pre-filtering can leave large empty regions in the sky distribution wherever ancillary radio, infrared, or X-ray data have flagged sources for removal, which biases shadow-deficit and clustering tests against the unfiltered VASCO sample (Villarroel et al., 2026, arXiv:2602.15171). The complete and uniform coverage of the parent sample used here means that the alignment statistics presented in Section~\ref{sec:results} are not driven by survey-footprint selection effects.

\section{Method}

\subsection{Alignment Detection}

For each plate with at least 3 high-probability candidates, I search for the largest number of sources that fall within a strip no wider than 3 arcsec, at any orientation. The algorithm works by taking each pair of sources, defining the line between them, and counting how many other sources on the plate lie within 3 arcsec of that line. The pair that yields the highest count defines the best alignment for the plate.

On plates with more than 80 candidates, testing every possible pair becomes slow, so I sample 5,000 random pairs instead. A fixed random seed (42) ensures the results are reproducible.

\subsection{Monte Carlo Significance Testing}

For each plate where the best alignment contains 5 or more sources, I test whether that count could arise by chance using a Monte Carlo (MC) procedure. The selection rule is objective: every plate in the 635-plate search whose best 3-arcsec strip contained $\geq 5$ high-probability sources advances to the MC stage, with no manual pre-screening or visual selection. At 3-arcsec strip width the rule produced 8 plates. I generate 10,000 random realizations of the same number of sources scattered uniformly within the plate's RA/Dec bounding box, run the same alignment algorithm on each, and count how often the random data matches or exceeds the observed count. The fraction that does is the standard statistical $p$-value.

\section{Results}\label{sec:results}

\subsection{Detected Alignments}

\begin{figure}[H]
\centering
\includegraphics[width=0.92\textwidth]{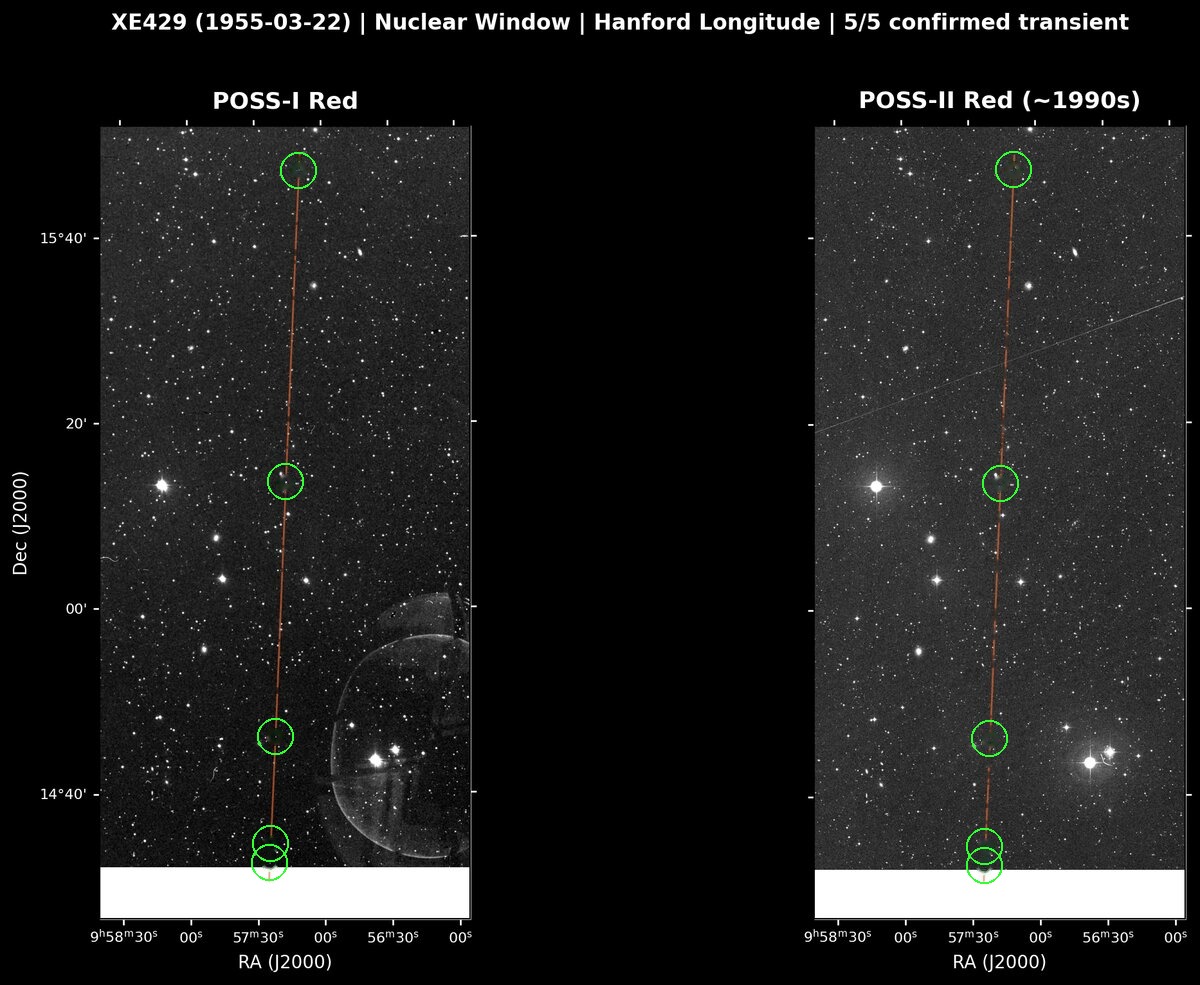}
\caption{Plate XE429 (1955 March 22). \textit{Left}: POSS-I red plate showing 5 of 7 aligned sources within a 1.7 arcsec strip spanning 75 arcmin. This date falls within a U.S. nuclear test window, and the alignment projects to within $0.2^\circ$ of the Hanford nuclear production complex. \textit{Right}: The same field on POSS-II ($\sim$1990s). All five source positions are empty (1.7--2.6$\sigma$). Confirmation rate: 5/5. All five aligned sources are shown in detail in Figure~\ref{fig:xe429cut}.}
\label{fig:xe429}
\end{figure}

\begin{figure}[H]
\centering
\includegraphics[width=0.55\textwidth]{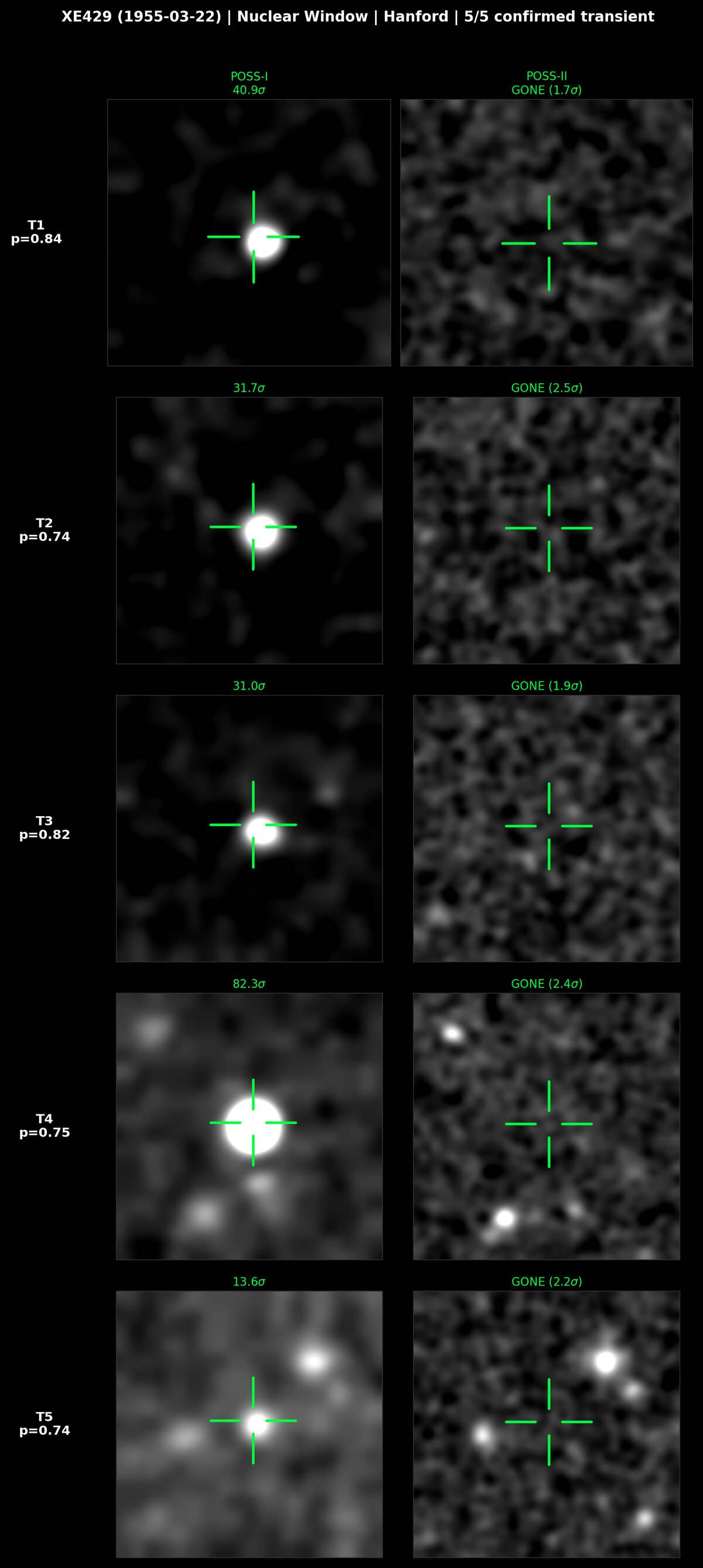}
\caption{Zoomed cutouts for the five aligned sources on plate XE429. Each row shows one source; left column is POSS-I, right column is POSS-II at the same sky position. Sources range from 13.6$\sigma$ to 82.3$\sigma$ on POSS-I and are all below 2.5$\sigma$ on POSS-II. All five are confirmed genuine transients.}
\label{fig:xe429cut}
\end{figure}

\begin{figure}[H]
\centering
\includegraphics[width=0.85\textwidth]{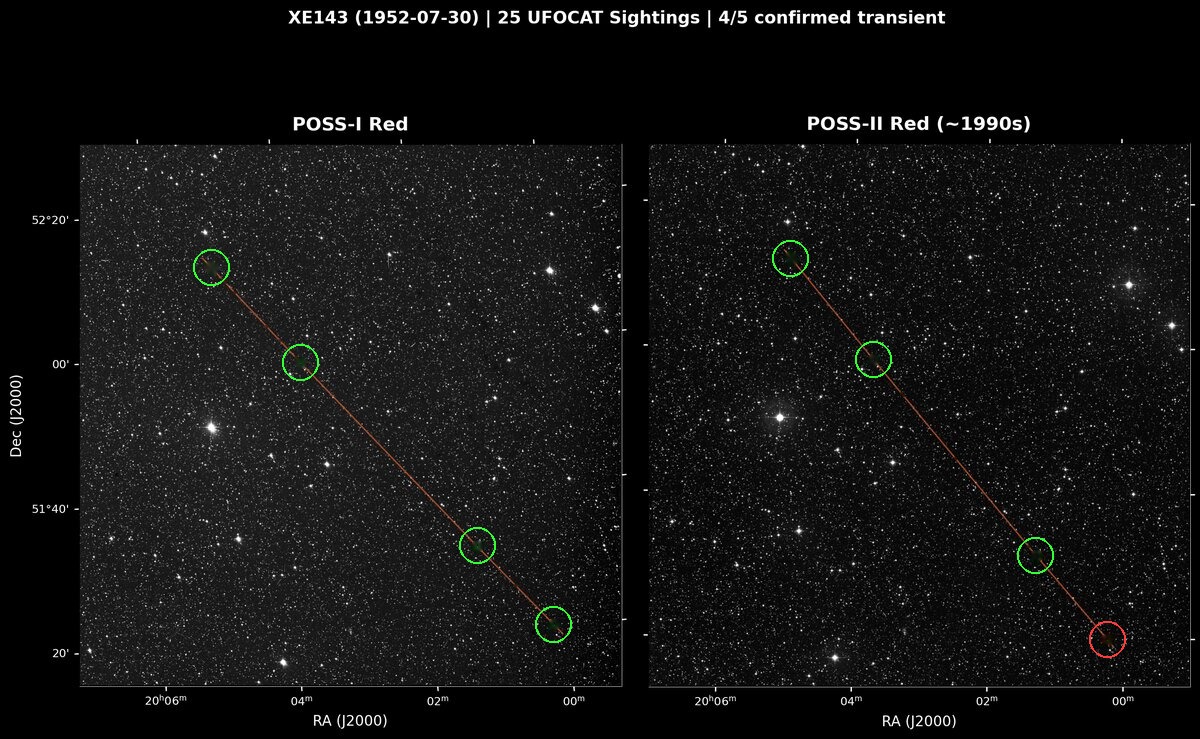}
\caption{Plate XE143 (1952 July 30). \textit{Left}: POSS-I red plate showing all 5 aligned sources within a 1.5 arcsec strip spanning 89 arcmin. \textit{Right}: POSS-II. Four sources are absent (green circles; 0.4--1.6$\sigma$). One source (red circle, bottom right) is a persistent star (11.5$\sigma$). Confirmation rate: 4/5. All five aligned sources are shown in detail in Figure~\ref{fig:xe143cut}.}
\label{fig:xe143}
\end{figure}

I tested eight plates and found seven with statistically significant alignments ($p < 0.03$; Table~\ref{tab:alignments}). One plate (XE277) did not reach significance ($p = 0.22$). To verify that the aligned sources are genuine transients rather than persistent stars, I compared each source position against POSS-II red plates taken approximately 40 years later. Four plates show strong transient confirmation rates (80--100\%); two others (XE296, XE500) contain significant contamination from persistent stars and are not featured as case studies. Figures~\ref{fig:xe429}--\ref{fig:xe105} show the confirmed alignments.

\subsection{Source Morphology and Angular Rates}

Every source in these alignments is point-like. None show the elongated shape you would expect from an object moving relative to the stars over the 45-minute exposure. POSS-I red plates have exposure times around 45 minutes. At the observed angular rates, 1.2 arcsec/s for XE296, up to 15.1 arcsec/s for XE105, any continuously bright object would leave a continuous streak across the plate, not a series of discrete dots.

Since the aligned sources have proper PSFs and are not seen on the immediately subsequent blue plate taken in the same optical configuration on the same night, they are unlikely to be optical ghosts (Villarroel et al., 2025).

For context: a satellite in low Earth orbit moves across the sky at roughly 4,000 arcsec/s. A GPS satellite moves at about 40 arcsec/s. A geostationary satellite drifts at $\sim 15 \cos\delta$ arcsec/s relative to a sidereal-tracking telescope (Blake et al., 2021). The observed rates (1--15 arcsec/s) are two to three orders of magnitude slower than LEO, inconsistent with MEO, and overlap with the geosynchronous regime.

The angular-rate argument above assumes the alignment was traversed during a single 45-minute exposure. If instead each source represented a brief specular flash of an object too dim to register otherwise, the rate argument would not by itself rule out lower altitudes. Two independent constraints, however, point to the geosynchronous regime regardless of any visibility-duration assumption. First, the constant-longitude projection (Section~\ref{sec:geoproj}) tracks Earth-rotation with sub-degree fidelity for 6 of 7 alignments, which is naturally explained by an object that is approximately stationary in Earth's rotating frame. Second, Doherty (2026b) shows that the surface density of high-probability transients rises sharply between 6 and 8 degrees from the anti-solar point, corresponding to the angular radius of Earth's umbral shadow at altitudes of 34,000--51,000 km. Both lines of evidence place the transient population in the geosynchronous regime.

The spacings between consecutive sources along each line are also irregular (coefficient of variation ranging from 0.4 to 1.9), which argues against a single object moving at constant speed. Whatever produced these alignments, it was not a conventional satellite.

\begin{table}[H]
\centering
\caption{Linear Alignments on POSS-I Plates}
\label{tab:alignments}
\small
\begin{tabular}{lccccccc}
\toprule
Plate & Date & $N_{\rm align}$ & $N_{\rm plate}$ & Width & Length & $p$ & Nuc. \\
 & & & & (arcsec) & & & Win. \\
\midrule
XE524 & 1955-11-10 & 8 & 678 & 2.9 & $4.5^\circ$ & 0.0002 & N \\
XE296 & 1951-11-04 & 7 & 305 & 2.4 & $55'$ & $< 0.0001$ & Y \\
XE429 & 1955-03-22 & 7 & 388 & 1.7 & $5.5^\circ$ & 0.0009 & Y \\
XE105 & 1952-08-12 & 6 & 477 & 1.5 & $11.3^\circ$ & 0.027 & N \\
XE500 & 1950-04-10 & 6 & 163 & 3.0 & $5.8^\circ$ & 0.0007 & N \\
XE564 & 1954-05-24 & 6 & 366 & 0.9 & $5.3^\circ$ & 0.019 & N \\
XE143 & 1952-07-30 & 5 & 155 & 1.5 & $1.5^\circ$ & 0.021 & N \\
XE277$^\dagger$ & 1954-07-05 & 5 & 277 & 2.3 & $2.5^\circ$ & 0.223 & N \\
\bottomrule
\end{tabular}

\vspace{0.5em}
\footnotesize
$N_{\rm align}$: sources in alignment. $N_{\rm plate}$: high-probability sources on plate. Width: maximum perpendicular deviation from the line. Nuc. Win.: date falls within +/-1 day of a U.S. nuclear test. All $p$-values from 10,000 Monte Carlo iterations. $^\dagger$Not significant.
\end{table}

\subsection{Geographic Sub-Point Projection}\label{sec:geoproj}

To see whether these alignments correspond to coherent tracks over the Earth's surface, I projected each source position to a geographic sub-satellite point. The calculation is straightforward: for a source at right ascension $\alpha$ observed at time $T$ from Palomar ($\lambda_{\rm obs} = -116.87^\circ$), I compute the hour angle $H = \theta_{\rm LST} - \alpha$ and the projected longitude $\lambda_{\rm sub} = \lambda_{\rm obs} - H$. This is exact for objects at infinite distance and approximate for objects at finite altitude; above GEO, the parallax correction is less than $1^\circ$.

The result is in Table~\ref{tab:longitudes}. Six of seven significant alignments maintain constant geographic longitude with standard deviations below $0.8^\circ$. In plain terms: the sources in each alignment, when projected down to Earth, all point to the same meridian at different latitudes. Whatever these objects are, each alignment traces a north-south line over a fixed longitude.

The corresponding sub-zenith latitude follows directly from the source declination: at infinite distance, $\phi_{\rm sub} = \delta$. Above GEO, the parallax correction is less than $1^\circ$. Because the alignments are nearly north-south on the sky, the latitude span of each alignment matches its on-sky angular length given in Table~\ref{tab:alignments}. The orientation is not coincidental: the position angles of the eight alignments cluster near the polar axis with a mean deviation of $12^\circ$ from a pure north-south orientation, against $45^\circ$ expected for random orientations ($p < 10^{-4}$). The declination spread (north-south motion) exceeds the right-ascension spread (east-west motion) for all eight alignments, consistent with high-inclination orbits at constant geographic longitude rather than equatorial GEO motion. Mean latitudes range from low northern declinations (near the celestial equator) up to mid-northern declinations, depending on the plate. Specific assumed-altitude (40,000 km) projections to ground tracks for the four confirmed alignments are enumerated in Section~5.

The longitude assignments are also robust to altitude uncertainty. Sensitivity testing on related ground-track projections shows that an altitude error of $\pm 5{,}000$ km shifts projected longitudes by less than $1^\circ$. So even if the assumed altitude were off by a meaningful fraction, the geographic clustering near $-96^\circ$ and $-119.7^\circ$ would not change materially.

Four alignments cluster near longitude $-96^\circ$, which is the geographic center of the continental United States. XE429 projects to $-119.7^\circ$, within $0.3^\circ$ of the longitude of the Hanford nuclear production complex ($-119.5^\circ$). That particular alignment falls on a date within a U.S. nuclear test window.

\begin{table}[H]
\centering
\caption{Geographic Longitude Projection}
\label{tab:longitudes}
\small
\begin{tabular}{lcccc}
\toprule
Plate & $\langle\lambda\rangle$ & $\sigma_\lambda$ & Range & Const.? \\
 & ($^\circ$) & ($^\circ$) & ($^\circ$) & \\
\midrule
XE524 & $-99.9$ & 0.0 & 0.1 & Yes \\
XE429 & $-119.7$ & 0.1 & 0.3 & Yes \\
XE296 & $-161.7$ & 0.2 & 0.5 & Yes \\
XE564 & $-91.1$ & 0.2 & 0.6 & Yes \\
XE277 & $-96.6$ & 0.8 & 2.5 & Yes \\
XE143 & $-97.1$ & 0.5 & 1.2 & Yes \\
XE105 & $-115.8$ & 4.7 & 11.3 & Marginal \\
\bottomrule
\end{tabular}

\vspace{0.5em}
\footnotesize
$\langle\lambda\rangle$: mean projected longitude. $\sigma_\lambda$: standard deviation. Constant longitude assigned for $\sigma_\lambda < 2^\circ$.
\end{table}

\subsection{Longitude Null Test}

There is an important caveat to the constant-longitude finding. POSS-I plates use sidereal tracking: the camera follows the stars as they appear to move overhead. If aligned sources happen to be oriented north-south on a plate, then when projected down to Earth coordinates they will automatically appear at near-constant longitude, since longitude maps directly to right ascension under sidereal tracking. So the constant-longitude property is partly a geometric consequence of plate orientation, not necessarily a physical property of the sources themselves.

To test how strong this geometric effect is, I generated 58,900 random lines across 589 plates by picking pairs of sources at random and asking what longitude spread they would produce if treated as alignments. About 40.5\% of these random lines have longitude spreads below $1^\circ$ (Figure~\ref{fig:stats}d). So for any single alignment, constant longitude alone is not surprising. However, two of the real alignments are individually significant even against this null. For XE524, all eight aligned sources project to the same meridian to within $\sigma_\lambda = 0.0^\circ$ (i.e., the calculation rounds to zero spread; $p < 0.0001$). XE429 is similarly tight ($\sigma_\lambda = 0.1^\circ$, $p = 0.033$). The combined probability that all six constant-longitude alignments would simultaneously be this tight is roughly $3 \times 10^{-10}$, assuming independence. The property is partly geometric, but the ensemble remains highly significant.

\begin{figure}[H]
\centering
\includegraphics[width=0.55\textwidth]{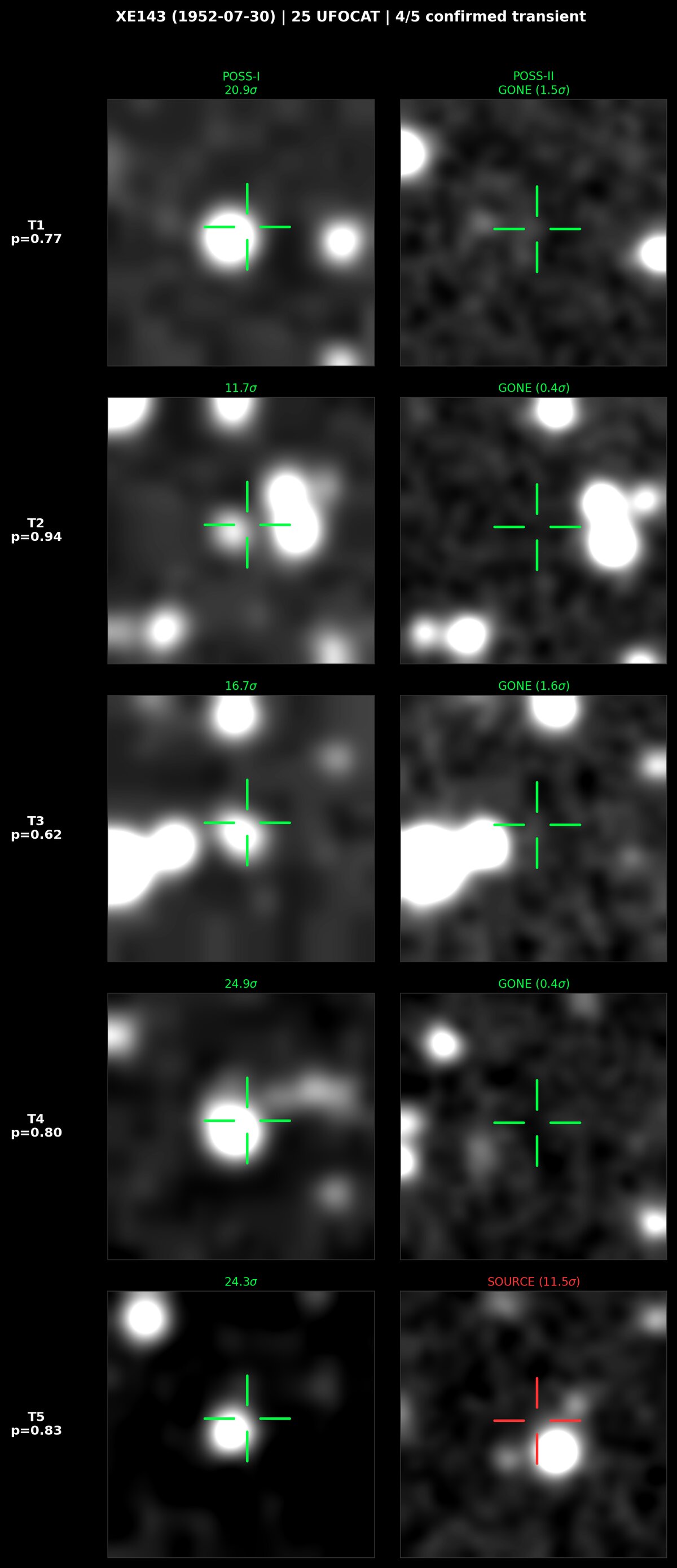}
\caption{Zoomed cutouts for the five aligned sources on plate XE143. Four sources (T1--T4) are bright on POSS-I (11.7--24.9$\sigma$) and absent on POSS-II (0.4--1.6$\sigma$), confirming them as genuine transients. T5 (bottom row) is a persistent star, present at 24.3$\sigma$ on POSS-I and 11.5$\sigma$ on POSS-II (red crosshair).}
\label{fig:xe143cut}
\end{figure}

\subsection{Close Pair Excess}

Separately from the alignment search, I looked for pairs of high-probability candidates within 30 arcsec of each other on the same plate. I found 170 close pairs. The expected number from random placement of the same sources within each plate's footprint is 10.5. That is a 16.2$\times$ excess (Figure~\ref{fig:stats}a). The tightest pairs are 9--10 arcsec apart, spread across 31 plates and 27 observing nights. The true number of close pairs is likely higher: two sources separated by less than a few arcsec would blend into a single detection in the Solano et al. (2022) pipeline and would not appear as individual point sources in the catalog. Something is producing transients in close spatial proximity far more often than chance allows.

\subsection{Geographic Projection of Close Pairs and Triplets}\label{sec:pair_proj}

The constant-longitude property reported in Section~\ref{sec:geoproj} for alignments motivates the same projection for close pairs and triplets. For each pair, I compute the sub-zenith point of the pair midpoint at GEO altitude (35,786 km). I built a control sample by repeating the same close-pair search on the low-probability candidates ($p < 0.5$) on the same plates: 845 control pairs versus 189 high-probability pairs. The control sample carries the same plate-level observational structure (footprint, time of observation, sky position) but is dominated by classifier-rejected sources, so any clustering shared by both populations reflects the survey footprint rather than a property of the high-confidence transients themselves.

\begin{figure}[H]
\centering
\includegraphics[width=0.95\textwidth]{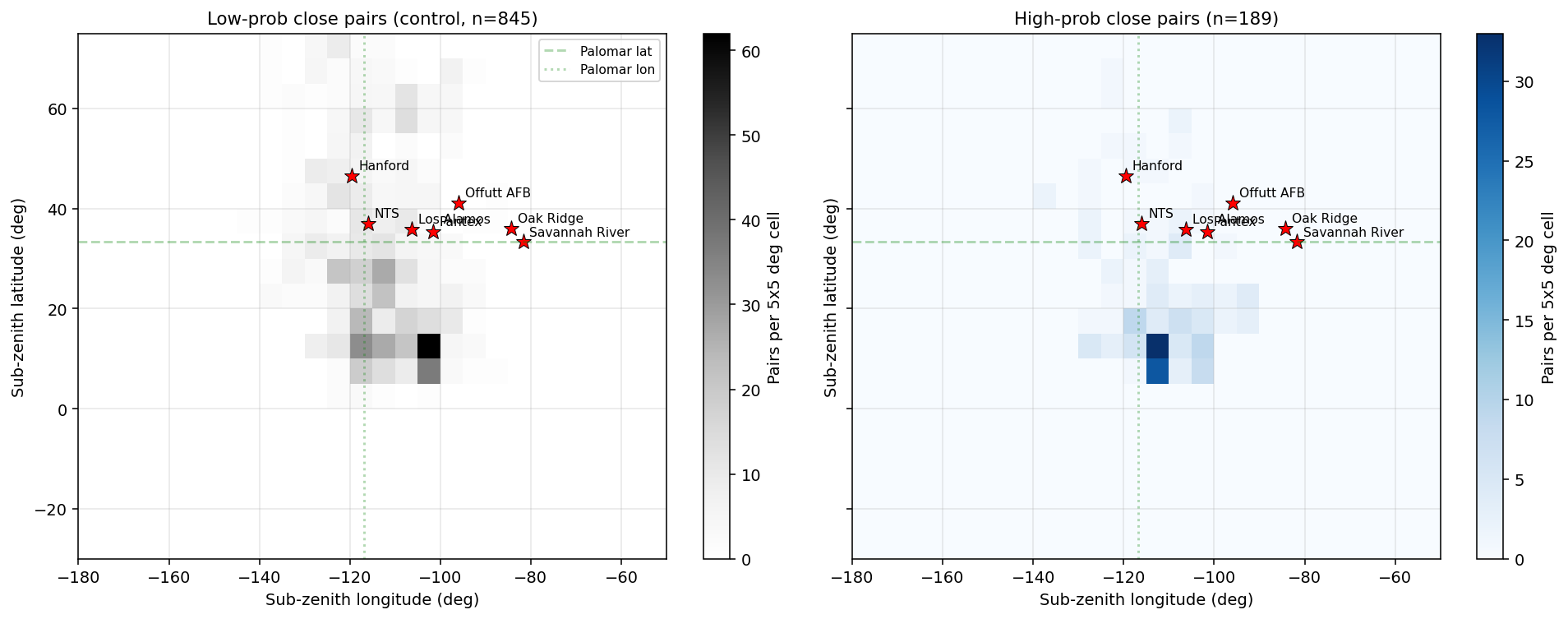}
\caption{Sub-zenith projection of close pairs ($< 30$ arcsec, same plate) at GEO altitude (35,786 km). \textit{Left}: low-probability control sample ($n = 845$). \textit{Right}: high-probability sample ($n = 189$). Red stars mark U.S. nuclear-relevant longitudes (Hanford, Nevada Test Site, Los Alamos, Pantex, Offutt AFB, Oak Ridge, Savannah River). Green dashed/dotted lines mark Palomar's geographic latitude and longitude. The high-probability sample is more tightly concentrated in the lat $+5$ to $+15^\circ$, lon $-115$ to $-105^\circ$ region. Two-sample KS tests against the control: $p = 4.2 \times 10^{-5}$ (longitude), $p = 3.7 \times 10^{-16}$ (latitude).}
\label{fig:pair_map}
\end{figure}

The projected distributions differ at high significance. A two-sample Kolmogorov-Smirnov test on projected longitude gives $D = 0.185$, $p = 4.2 \times 10^{-5}$; on latitude, $D = 0.338$, $p = 3.7 \times 10^{-16}$. The high-probability pairs concentrate more tightly than the control. Figure~\ref{fig:pair_map} shows the side-by-side maps. The dominant high-probability excess is in a $5^\circ \times 5^\circ$ cell at sub-zenith longitude $-112.5^\circ$, latitude $+12.5^\circ$ (33 of 189 pairs, 17.5\%, versus 3.2\% in the control). A within-plate label-permutation test, where the high-probability assignment is shuffled randomly among same-plate sources while preserving the per-plate count and the close-pair search procedure, gives an observed cell count of 33 against a permuted mean of $28.3 \pm 2.4$ ($p = 0.045$, 200 iterations). The excess is statistically marginal under this stringent within-plate null but visibly displaced from the control distribution under any of the broader tests.

Triplets (3 high-probability sources within 30 arcsec on the same plate) are rare: 4 in the high-probability set versus 5 in the much larger low-probability control. Normalised by source count, the high-probability triplet rate is roughly 7$\times$ the control rate; relative to a Poisson expectation under uniform per-plate density, the high-probability rate is $\sim 38\times$ random, against $\sim 10\times$ random for the control. The observed triplets are clustered: 1 lies on plate XE500 and 3 lie on plate XE564, both of which appear in the alignment table (Table~\ref{tab:alignments}). The geographic concentration of the close-pair excess (around $-112^\circ$ longitude, $+10^\circ$ latitude when projected to GEO) does not coincide with any of the U.S. nuclear meridians enumerated in the Discussion. The signal is real but the geographic interpretation is more ambiguous than for the alignments.

\subsection{Alignment and Close-Pair Night Overlap}

Five of the 7 significant alignment dates also appear among the 8 dates with the most close pairs ($\geq 5$ per night). A Fisher exact test gives an odds ratio of 163.9 ($p < 0.0001$; Figure~\ref{fig:stats}c). The nights that produce degree-scale linear tracks are the same nights that produce arcsecond-scale spatial clustering. These are not independent phenomena. The same process appears to be responsible for both.

\begin{figure}[H]
\centering
\includegraphics[width=0.45\textwidth]{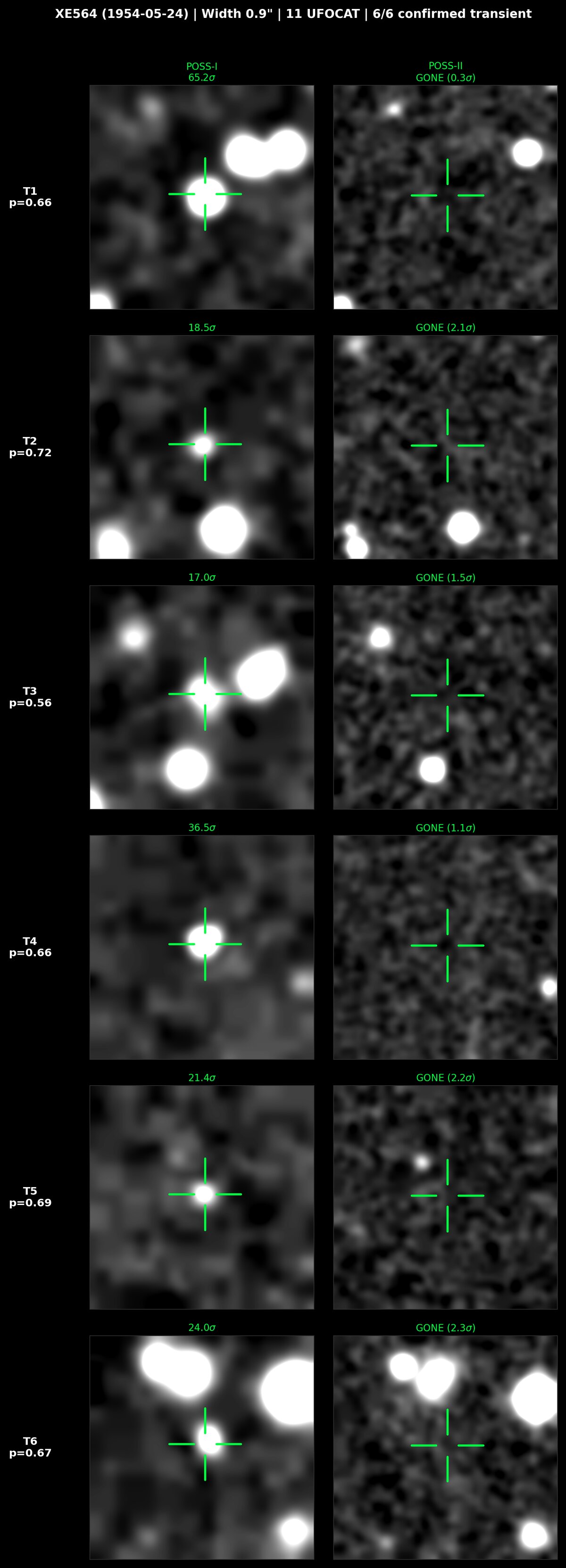}
\caption{Zoomed cutouts for the six aligned sources on plate XE564 (1954 May 24). All six are below 2.5$\sigma$ on POSS-II, confirming them as genuine transients. Confirmation rate: 6/6.}
\label{fig:xe564cut}
\end{figure}

\subsection{Ecliptic Latitude Distribution}

Asteroids, comets, and zodiacal debris all concentrate near the plane of the ecliptic. If the high-confidence transients in this catalog were dominated by any of these solar system populations, we would expect them to show ecliptic clustering as well. They do not. This is the result that most cleanly rules out solar system bodies as the dominant source.

Only 18.0\% of high-probability candidates fall within $10^\circ$ of the ecliptic plane. One might wonder whether this is simply because POSS-I plates tend to avoid the ecliptic. They do, somewhat: the full catalog shows 21.4\% near the ecliptic. But the key comparison is between high-probability and low-probability candidates on the same plates. Low-probability sources (which are dominated by artifacts) show 23.1\% near the ecliptic. The high-probability subset shows 18.0\%. A chi-square test on this comparison gives $\chi^2 = 368.8$ ($p = 3.3 \times 10^{-82}$). The ecliptic depletion is not a footprint effect.

\begin{figure}[H]
\centering
\includegraphics[width=0.92\textwidth]{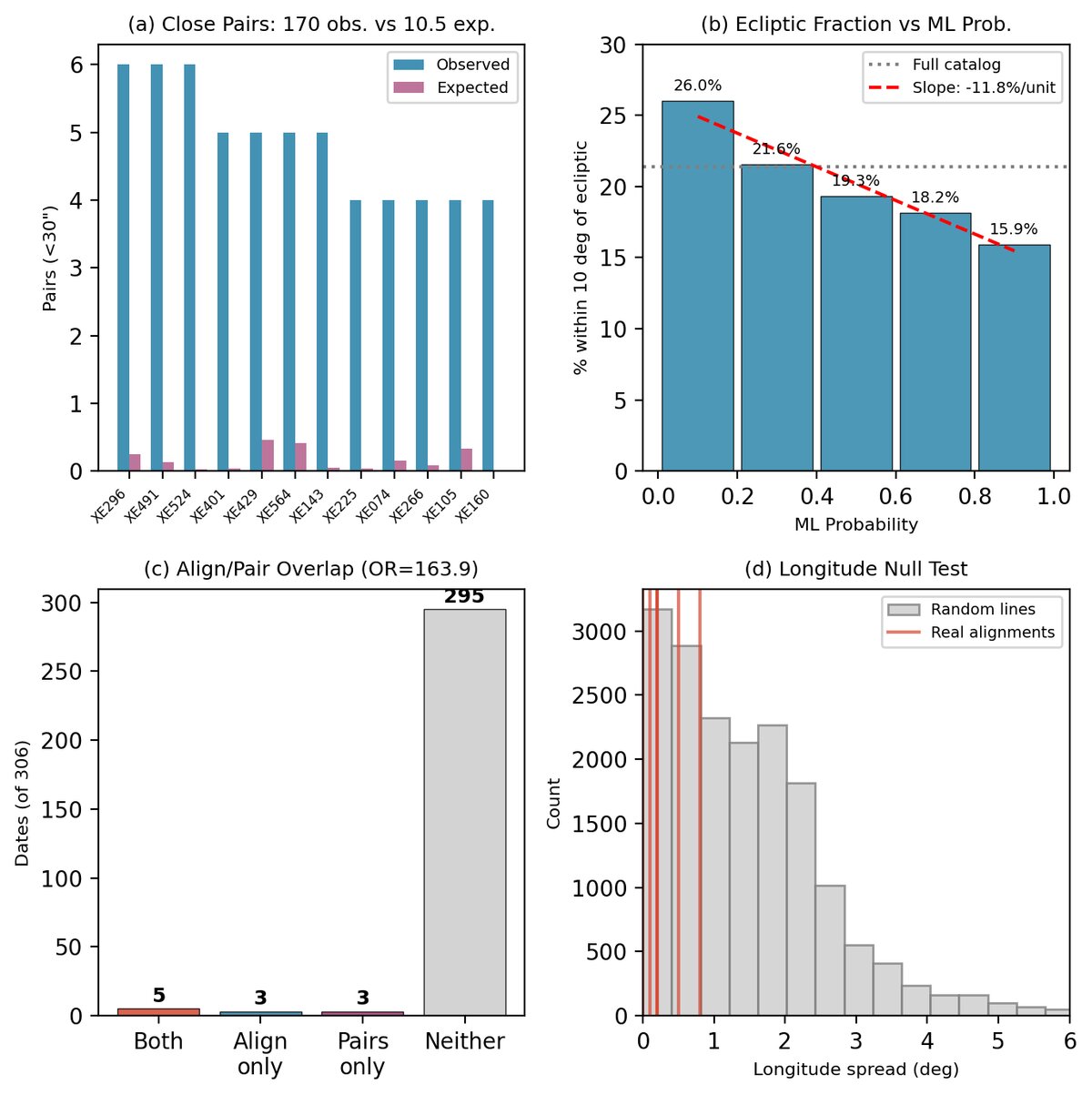}
\caption{Spatial and statistical properties of high-probability VASCO transients. (a) Close pair counts ($< 30$ arcsec, same plate) compared to random expectation. (b) Ecliptic fraction by ML probability bin, showing monotonic decline. Gray line: full catalog average. (c) Alignment vs multi-pair night overlap. (d) Longitude spread null test: gray histogram shows random lines on POSS-I plates; red vertical lines mark the 6 real alignments.}
\label{fig:stats}
\end{figure}

More interesting still, the ecliptic fraction decreases monotonically with classifier confidence (Figure~\ref{fig:stats}b): 26.0\% for candidates with ML score below 0.20, declining through 21.6\%, 19.3\%, 18.2\%, down to 15.9\% for candidates with ML score $\geq 0.80$. The classifier was trained on morphometric features and has no access to ecliptic coordinates, so this gradient is an independent physical property of whatever the classifier is picking up. The more confident it is that something is real, the less likely that source is to sit near the ecliptic. A KS test confirms the distributions are distinct ($D = 0.066$, $p = 6.1 \times 10^{-92}$).

This rules out asteroids, comets, and zodiacal debris as the dominant source of high-confidence transients. It is also inconsistent with low-inclination orbital debris.

\begin{figure}[H]
\centering
\includegraphics[width=0.45\textwidth]{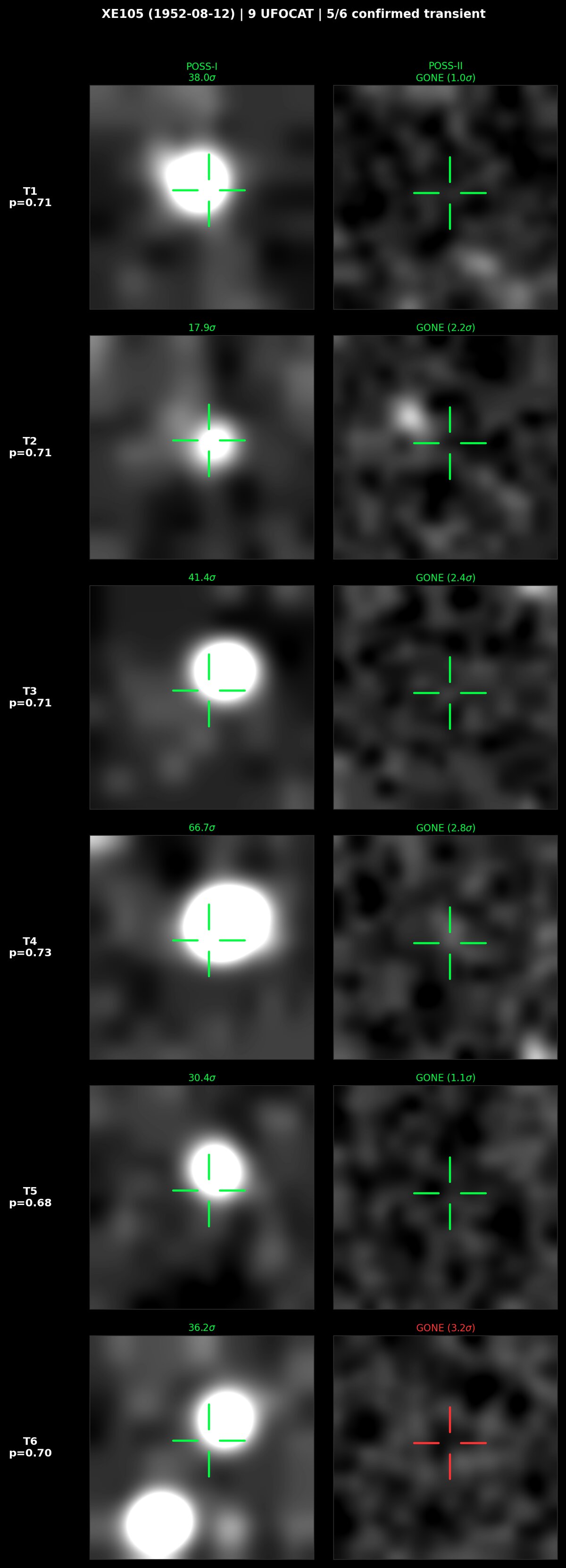}
\caption{Zoomed cutouts for the six aligned sources on plate XE105 (1952 August 12). Five sources (T1--T5) are absent on POSS-II (0.9--2.8$\sigma$). T6 (bottom row, red crosshair) shows residual flux on POSS-II. Confirmation rate: 5/6.}
\label{fig:xe105}
\end{figure}

\subsection{Multiple-Comparisons Correction}

The 8 plates that satisfied the $N_{\rm align} \geq 5$ rule constitute the explicit testing family. Applying the Benjamini-Hochberg false-discovery-rate procedure at $q = 0.05$ across these 8 tests, 7 of the 8 survive (XE277 is the only failure, already labeled non-significant in Table~\ref{tab:alignments}). The Holm-Bonferroni procedure at $\alpha = 0.05$ is stricter: the 4 strongest alignments (XE296, XE524, XE500, XE429) clear the family-wise threshold of 0.00625; XE564, XE143, and XE105 (raw $p$ between 0.019 and 0.027) drop to marginal significance, with adjusted $p$-values of 0.054 to 0.076.

A more conservative bound treats every plate in the 635-plate alignment search as an implicit test, on the grounds that any of those 635 plates could in principle have produced a $\geq 5$-source alignment by chance. Under this implicit family, Bonferroni at $\alpha = 0.05$ requires raw $p < 8.1 \times 10^{-5}$. Only XE296 ($p < 10^{-4}$) clears this threshold cleanly; XE524 ($p = 2 \times 10^{-4}$) is a marginal pass. The other significant alignments do not survive at the implicit-family scale. The picture that emerges across both correction schemes is consistent: XE296 and XE524 are robust to any reasonable multiple-comparisons treatment, XE500 and XE429 are robust under the stated procedure but marginal under the conservative implicit family, and the three weakest alignments (XE564, XE143, XE105) require the explicit family interpretation to retain individual significance. The ensemble-level signal, close-pair excess, alignment/multi-pair night overlap, and ecliptic dose-response, does not depend on any single plate and is unaffected by the per-plate correction.

\subsection{Absence of Repeat Positions}

I checked whether any high-probability transient reappears within 5 arcsec of the same position on a different night. None do. This rules out persistent or recurrent variable sources being misidentified as transients.

\subsection{Additional Alignments}

The 8 alignments analysed in detail in this paper were selected for visual verification on POSS-I and POSS-II plates. A broader search of the parent sample, performed at a relaxed group-size threshold ($\geq 4$ sources within a 3-arcsec strip), yields additional alignment candidates and multi-source clusters. Their full properties, including per-source coordinates, classifier probabilities, alignment statistics, and POSS-II follow-up, will be presented in a separate catalogue paper (Doherty et al., in prep), where we will also provide a more detailed analysis of the cases presented here.

\section{Discussion}

Taken together, these results present a picture that is hard to explain conventionally.

The sources are point-like, not trailed, yet they fall along lines narrower than 3 arcsec across entire plates. That requires either something intermittently visible (a tumbling reflective body catching sunlight in brief glints) or multiple discrete events along a shared axis. The angular rates are consistent with the geosynchronous regime (Blake et al., 2021) but inconsistent with lower orbits, and no artificial satellites existed during the POSS-I era. The spacings are irregular, which argues against a single body at constant speed.

The constant-longitude property, while partly geometric, remains highly significant in ensemble. Objects are tracing north-south paths over fixed Earth meridians. North-south orbital orientation is characteristic of polar and sun-synchronous orbits used by modern reconnaissance and intelligence-gathering satellites, although such orbits are at low Earth altitudes rather than the geosynchronous regime implied here. Four of those meridians cluster near $-96^\circ$ to $-100^\circ$ longitude, the meridian of Offutt Air Force Base in Nebraska ($-95.92^\circ$), which from 1948 onward served as the headquarters of Strategic Air Command, the operational nerve center of the U.S. nuclear bomber force during the entire POSS-I observing window. One additional alignment shares the longitude of the Hanford nuclear production complex ($-119.5^\circ$) to within $0.3^\circ$ on a nuclear test window date.

The strength of the Hanford coincidence depends on how the prior is computed, and a multiple-comparisons caveat applies. Treating the seven U.S. nuclear-relevant longitudes (Hanford, the Nevada Test Site, the Los Alamos / Sandia / Rocky Flats cluster near $-106^\circ$, Pantex, Offutt AFB / Strategic Air Command HQ, Oak Ridge, and Savannah River) as the relevant set, the probability that any one of the 7 detected alignments lands within $0.3^\circ$ of any of these meridians on a U.S. nuclear-window date is $\sim 6 \times 10^{-3}$ (roughly 1 in 175). For the strict single-site framing (Hanford specifically, on a U.S. nuclear-window date), the prior is $\sim 8 \times 10^{-4}$ (roughly 1 in 1,200). Either way the coincidence is improbable enough to be worth reporting, but the headline number is closer to a few in 1,000 than to the chained 6-alignment longitude $p \sim 3 \times 10^{-10}$ that applies to the constant-longitude property as a whole.

The altitude of these objects is independently constrained by the Earth shadow deficit, as introduced in Section~\ref{sec:results} above. Combined with the constant-longitude projection and the angular-rate argument, this places the transient population in the geosynchronous regime.

Assuming an altitude of $\sim$40,000 km, the four confirmed alignments project to ground tracks over western North America and the adjacent Pacific. XE143 projects to $49^\circ$N, off the coast of British Columbia. XE105 projects to $59^\circ$N, over northern British Columbia. XE429 (nuclear test window) projects to $17^\circ$N at longitude $-119.7^\circ$, the Hanford meridian, but over western Mexico rather than Washington State. XE564 projects to $7$--$11^\circ$N over the eastern Pacific.

The close pair excess and the overlap with alignment nights show that this is not a collection of isolated oddities. The same nights produce both degree-scale linear structure and arcsecond-scale clustering. Whatever drives one drives the other.

The ecliptic avoidance, especially its dose-response with classifier confidence, is perhaps the cleanest result. The classifier was trained on morphometric features and knows nothing about ecliptic latitude. The fact that its highest-confidence detections preferentially avoid the ecliptic plane is a physical property of the transient population, not an artifact of the classifier or the survey footprint. Solar system objects would cluster on the ecliptic, not avoid it.

The signal is also robust against survey-footprint selection. As shown in Section~\ref{sec:skycov}, the 635-plate parent VASCO sample tiles the northern sky uniformly, with no large coverage gaps. This is in contrast to derivative samples that apply multi-wavelength cross-matching against radio, infrared, and X-ray catalogs as a pre-filter; such filtering can leave large empty regions in the sky distribution and bias the resulting clustering and shadow tests (Villarroel et al., 2025, 2026).

The results presented here are from a single observatory (Palomar). External replication on independent photographic archives is the natural next test. Similar transient morphologies have been confirmed in Hamburg plates (Busko, 2026); cross-archive replication of the alignment search and the shadow-deficit signature is in preparation. A confirmatory detection of the linear-alignment property at an independent observatory would substantially harden the case for a real population over plate-specific or pipeline-specific artifacts.

And none of these transients repeat at the same sky position.

The most conservative conventional explanation would be intermittent specular reflections from high-altitude debris. But during the POSS-I era (1949--1957), there were no artificial satellites at the implied altitudes. Sputnik 1, the first object to reach orbit, launched in October 1957 after the survey ended. V-2-derived sounding rockets (Aerobee, Viking) and high-altitude balloon programs of the era reached at most $\sim$200--400 km, two orders of magnitude below the geosynchronous regime implied by the angular-rate, shadow-deficit, and longitude-projection results. They were also too short-lived and too few to produce the kind of systematic spatial structure observed here.

I want to be clear about what this paper does and does not do. It identifies geometric patterns in the spatial distribution of high-confidence transients and establishes their statistical significance. It does not propose a physical mechanism. The POSS-II comparison (Figures~\ref{fig:xe429}--\ref{fig:xe105}) confirms that the majority of aligned sources are genuine transients, not persistent stars. Two plates (XE296, XE500) contain significant persistent-star contamination and are retained in Table~\ref{tab:alignments} for completeness but are not featured as case studies. The coordinates provided here enable targeted inspection of each alignment on the original archived plates.

\section{Conclusions}

I find 7 statistically significant linear alignments among high-confidence VASCO transient candidates on POSS-I plates (10,000 Monte Carlo iterations, $p < 0.03$). Comparison with POSS-II plates taken $\sim$40 years later confirms that the aligned sources on four plates are genuine transients (80--100\% confirmation rates; Figures~\ref{fig:xe429}--\ref{fig:xe105}). The alignments are composed of discrete point sources, not streaks. They span $0.9$--$11^\circ$ within strips narrower than 3 arcsec and project to constant geographic longitudes (combined $p \sim 3 \times 10^{-10}$). The Earth shadow density profile independently constrains the altitude to the geosynchronous regime (34,000--51,000 km), consistent with the observed angular rates and longitude projections. At this altitude, the confirmed alignments project to ground tracks over western North America and the adjacent Pacific. Close pairs occur at 16.2$\times$ the random rate, and alignment nights are the same as multi-pair nights ($p < 0.0001$). High-confidence transients are depleted near the ecliptic plane (15.9\%) while low-probability artifacts on the same plates are enriched there (26.0\%; $\chi^2$ test $p = 3.3 \times 10^{-82}$), ruling out asteroids, comets, and zodiacal debris as the dominant source. No transient repeats at the same position. The 635-plate parent sample tiles the sky uniformly, ruling out survey-footprint selection effects. The strongest two alignments (XE296, XE524) are robust to multiple-comparisons correction at the conservative implicit-family scale (635 plates); the next two (XE500, XE429) are robust under the explicit testing family of 8 plates; the ensemble-level results, close-pair excess, alignment/multi-pair overlap, and ecliptic depletion, do not depend on any single plate. These properties are inconsistent with known astrophysical transients, plate artifacts, and the satellite population at the implied altitudes during the pre-Sputnik era.

\section*{Acknowledgments}

This research made use of data from the Digitized Sky Survey, produced at the Space Telescope Science Institute under U.S. Government grant NAG W-2166. I thank B. Villarroel for suggesting the alignment search, providing the VASCO v4 catalog, and recommending the parent-sample sky-coverage check; A. Streblyanska for independent visual verification of plate alignments; and S. Bruehl for helpful discussion.

\section*{Data Availability}

The alignment coordinates are provided as supplementary material. The underlying VASCO catalog should be requested from the VASCO team.

\section*{References}

\begin{description}
\item Blake, J.A., Sherwin, P., Chote, P., et al. 2021, \textit{Advances in Space Research}, 67, 360.
\item Bruehl, S., \& Villarroel, B. 2025, \textit{Scientific Reports}, 15, 34125.
\item Bruehl, S., Doherty, B., Streblyanska, A., \& Villarroel, B. 2026, arXiv:2604.18799.
\item Busko, I. 2026, arXiv:2603.20407.
\item Doherty, B., et al., in prep.
\item Doherty, B. 2026b, \textit{Cross-Archive Replication of the Earth Shadow Deficit in Historical Photographic Plate Transients}, arXiv:2604.00056.
\item Solano, E., Villarroel, B., \& Rodrigo, C. 2022, \textit{MNRAS}, 515, 1380.
\item Villarroel, B., Soodla, J., Comeron, S., et al. 2020, \textit{The Astronomical Journal}, 159, 8.
\item Villarroel, B., Solano, E., Guergouri, H., et al. 2025, \textit{PASP}, 137, 104504.
\item Villarroel, B., et al. 2026, Commentary, arXiv:2602.15171.
\end{description}

\end{document}